\documentclass{article}
\usepackage{graphicx} 

\usepackage[utf8]{inputenc}
\usepackage[T1]{fontenc}
\usepackage{mathpazo}
\usepackage{emptypage}
\usepackage{mathtools}
\usepackage{setspace}
\usepackage{scrlayer-scrpage}
\usepackage{natbib}
\AtBeginEnvironment{tabular}{\singlespacing}
\usepackage{relsize}
\usepackage{float}
\usepackage{amsmath}
\usepackage{accents}
\newcommand*{\dtt}[1]{%
  \accentset{\mbox{\large\bfseries .}}{#1}}
\usepackage{graphicx}
\usepackage{enumitem}
\usepackage{lipsum}
\usepackage{csquotes}
\usepackage{cancel}
\usepackage{pdfpages}
\providecommand{\keywords}[1]
{
  \small	
  \textbf{\textit{Keywords---}} #1
}
\usepackage{authblk}
\title{The generalized Hausman test for detecting non-normality in the latent variable distribution of the two-parameter IRT model}
\author[1]{Lucia Guastadisegni \thanks{lucia.guastadisegni2@unibo.it}}
\author[1]{Silvia Cagnone}
\author[2]{Irini Moustaki}
\author[3]{Vassilis Vasdekis}
\affil[1]{University of Bologna}
\affil[2]{London School of Economics and Political Science}
\affil[3]{Athens University of Economics and Business}
\date{}

\begin{document}

\maketitle
\begin{abstract}

This paper introduces  the generalized Hausman test as a novel method for detecting non-normality of the latent variable distribution of unidimensional Item Response Theory (IRT) models for binary data. The test utilizes the pairwise maximum likelihood estimator obtained for the parameters of the classical two-parameter IRT model, which assumes normality of the latent variable, and the quasi-maximum likelihood estimator obtained under a semi-nonparametric framework, allowing  for a more flexible distribution of the latent variable. The performance of the generalized Hausman test is evaluated through a simulation study and it is compared with the likelihood-ratio and the $M_2$ test statistics. Additionally, various information criteria are computed. The simulation results show that the generalized Hausman test outperforms the other tests under most conditions. 
However, the results obtained from the information criteria are somewhat contradictory under certain conditions, suggesting a need for further investigation and interpretation. 
\end{abstract}

\section{Introduction}
The aim of the paper is to propose a generalized Hausman type test for detecting deviation of the latent variable distribution from normality. The paper focuses on mixture of normals and skewed normal distributions. 

In many  fields of research, there is a need to study  theoretical constructs, such as abilities, quality of life, or business confidence, which cannot be directly observed and measured. To address this, latent variable models are employed, which measure the constructs of interest also known as latent variables or factors by analyzing the associations or 
correlations among observed variables also known as manifest variables or items (\citealp{bartholomew2011latent}).  
Latent variable models can accommodate both continuous and discrete variables, including binary and polytomous outcomes. In the case of binary or polytomous outcomes, Item Response Theory (IRT) models are commonly used. These models are a type of latent variable model where the observed outcomes are binary or polytomous, and the latent variables are assumed to be continuous (\citealp{van2013handbook}). The use of IRT models is particularly prevalent in social, psychological, and educational research, where they are employed to measure various constructs such as attitudes or abilities.

One of the standard assumptions of IRT models is that the latent variable(s) follow a normal distribution. However, assuming normality of the latent variable(s) when the true distribution has a different shape can lead to biased parameter estimates, especially with binary outcomes \citep{ma2010explicit}. Furthermore, assuming an incorrect distribution of the latent variable can lead to erroneous conclusions when conducting hypothesis testing \citep{guastadisegni2021use}. In the literature of the generalized linear latent variable models (GLLVM) (\citealp{bartholomew2011latent}, \citealp{skrondal2004generalized}) and  IRT models, several methods that assume a different form for the distribution of the latent variable have been proposed. \citet{montanari2010skew} introduce a skew-normal latent variable in the factor model, while \citet{cagnone2012factor} present a latent trait model where the factors are distributed as a finite mixture of multivariate gaussians. Within the GLLVM framework, \citet{ma2010explicit} propose a semi-parametric method, consistent for various types of manifest variables under different distributions of the latent variables and 
\citet{irincheeva2012generalized} consider the semi-non-parametric (SNP) approach, introduced by \citet{gallant1987semi}. This approach allows for more flexible smooth densities of the latent variables. The SNP method has been used also in unidimensional IRT model by \citet{woods2009item} and in multidimensional IRT model by \citet{monroe2014multidimensional}.
For binary responses, \citet{knott2007bootstrapping} estimate the distribution of the latent variable from the data, using the empirical histogram method combined with the bootstrap.
\citet{woods2006ramsay} proposes the so-called Ramsey curve IRT model, where the latent variables are splines based densities, that are linear combination of polynomial functions joint together at knots. This method implies a modification of the standard E-M algorithm.

 In the majority of cases, information criteria, such as the Akaike information criterion (AIC)  (\citealp{akaike1974new}) or the Bayesian information criterion (BIC) (\citealp{schwarz1978estimating}), are used to choose between a model where the latent variables are normally distributed and a model where the latent variables have a more complex shape (\citealp{woods2009item}, \citealp{irincheeva2012generalized}, \citealp{monroe2014multidimensional}). 
With continuous manifest variables, \citet{ma2010explicit} perform the Kolmogorov–Smirnov test on the distribution of the continuous responses to evaluate the normality of the latent variable. However, when the responses are categorical, detecting non-normality of the latent variables through a statistical test remains an open issue.

\citet{hausman1978specification} proposes a specification test to detect failure of the orthogonality assumption in regression analysis. Due to its simplicity, the Hausman test can be applied in various contexts to detect different types of model misspecification. This test is based on the comparison between two different estimators that are consistent when the model is correctly specified and one of them is also efficient. In presence of model misspecification, only the inefficient estimator is consistent. The efficiency assumption simplifies the computation of the covariance matrix of the difference between the two estimators. However, this matrix can fail to be positive definite under model misspecification or in presence of small sample sizes. Moreover, in some cases none of the two estimators considered are fully efficient.

The generalized version of the Hausman (GH) test, proposed by \citet{white1982maximum}, is a more flexible and robust alternative to the original Hausman test. Indeed, the generalized version allows for both estimators to be inefficient and to result from two different models.
Moreover, the covariance matrix of the difference of the two estimators is robust and always positive definite. 

In the IRT context, as far as we know, the Hausman test has been used only by \citet{ranger2020analyzing} to detect misspecification of the item characteristic functions and local dependencies among items. They highlight that this test has good performance in terms of Type I error rates for large sample sizes and power under most conditions. 
 In generalized linear mixed models (GLMM) for clustered data, a robust version of the Hausman test, similar to the one by \citet{white1982maximum}, has been proposed by \citet{bartolucci2017misspecification} when a discrete distribution for the random effects is assumed. The test can be also used to detect the possible correlation between random effects and cluster-specific covariates. With respect to the information criteria, they found that the robust Hausman test prefers more parsimonious models and it can detect the presence of endogeneity.

The objective of this work is to extend the GH test to detect non-normality of the latent variable distribution in unidimensional IRT models for binary data. To build the test, the estimators resulting from two different models are considered. 
The first model is a two-parameter logistic (2PL) unidimensional IRT model that assumes normality of the latent variable, while the second model is the unidimensional SNP-IRT model that assumes a more flexible distribution for the latent variable. The 2PL model is estimated using a pairwise maximum likelihood (PL) method (\citealp{katsikatsou2012pairwise},\citealp{vasdekis2012composite}), which is a composite likelihood method that uses information from bivariate-order margins (\citealp{lindsay1988composite}, \citealp{varin2008composite}). The SNP-IRT model is estimated using a quasi-maximum likelihood (quasi-ML) method (\citealp{white1982maximum}). The choice of these estimators and models is motivated by the following reasons. First, both methods are consistent when the latent variable is normally distributed. Moreover, the quasi-ML method for the SNP-IRT model is consistent also under
different distribution assumptions of the latent variable (\citealp{gallant1989seminonparametric}, \citealp{irincheeva2012generalized}). These conditions on the consistency of the parameter estimators are required to correctly apply the GH test (\citealp{white1982maximum}).
Second, the PL estimator and the quasi-ML estimator yield different values for the parameter variance. This implies that, also under normality of the latent variable distribution, the covariance matrix of the difference of the two estimators involved in the computation of the GH test is different from zero. The non-zero covariance matrix avoids numerical instability in the computation of the test.

The theoretical aspects of the models, estimators and matrices involved in the computation of the GH test are described.
Moreover, we carry out an extensive simulation study to evaluate the performance of the GH test both in terms of Type I error rates and empirical power. For the latter, we consider both mixture of normals and skew-normal distributions for the latent variable, with varying degrees of departure from the normal distribution. Additionally, we evaluate the asymptotic behavior of the test, in terms of both Type I error rates and power, using a very large sample size. The performance of the GH test is compared with the $M_2$ test, a statistic based on a quadratic form in marginal
residuals of order one and two, not affected by the problem of sparse data \citep{maydeu2005limited} and the likelihood-ratio (LR) test for nested models. In addition, some information criteria are computed. An application to real data is also presented.

The article is organized as follows. In Section 2, we describe the 2PL model and SNP-IRT models for binary data and the PL and quasi-ML estimators, respectively.  In Section 3, we present the GH test to detect non-normality of the latent variable distribution. In Section 4 we review the $M_2$ test and the LR test and in Section 5 the information criteria. In Section 6, we present a Monte Carlo simulation study and in Section 7 the results from a real data analysis. Finally, in Section 8, some concluding remarks are presented and discussed.
\section{The 2PL and SNP-IRT model for binary data}
Let us denote by $y_1,...,y_p$ a set of observed binary variables/items, by $n$ the number of individuals and by $z$ the latent variable with density function $h(z)$. 

According to the 2PL model, the response category probability for the $i$-th individual to the $j$-th item is modelled using a logistic model (measurement model) 
\begin{equation}
\begin{split}
     &P(y_{ij}=1|z_i)={\pi _{ij}}(z_i)=\frac{\exp{(\alpha_{0j}+\alpha_{1j}z_i)}}{1+\exp{(\alpha_{0j}+\alpha_{1j}z_i)}}, \hspace{0.8cm}j=1,\ldots,p
\end{split}
\label{mod1}   
\end{equation}
where $\alpha_{0j}$ is the item intercept and $\alpha_{1j}$ the item slope (factor loadings) and $h(z)=\phi(z)$, where $\phi(z)$ is the density of a standard normal. 
An extension of the IRT model is given by the SNP-IRT model that assumes the same response probability as (\ref{mod1}) but a semi-non-parametric parametrization of the latent variable as follows 
\begin{equation}
\begin{split}
    h(z_i)=P_L^2(z_i)\phi(z_i) \hspace{0.3cm}\text{and}\hspace{0.3cm}P_L(z_i)=\sum_{0\le l\le L}a_{i}z_i^{l},
\end{split}
\label{mod2.1}   
\end{equation}
where $a_{0},...,a_{L}$ are the real coefficients of the polynomial $P_L(z_i)$ and $L$ is the polynomial degree.

In order for $h(z)$ to be a density, the coefficients $a_0,...,a_L$ of $P_L(z)$ should be chosen such that $\int h(z)dz=1$.
For this purpose, \citet{gallant1989seminonparametric} use a proportionality constant $1/\int P_L(z)^2\phi(z) dz$ and fix the constant term of the polynomial equal to 1. Alternatively, \citet{woods2009item} and \citet{irincheeva2012generalized}  use the parametrization proposed by \citet{zhang2001linear}, that imposes
\begin{equation}
       1=\int_{R}P_L^2(z)\phi(z)dz=E\{P_L^2(w)\}=\textbf{a}'E(\boldsymbol{\tilde{w}}\boldsymbol{\tilde{w}}')\textbf{a}=\textbf{a}'A\textbf{a} \label{cond}
\end{equation}
with $w \sim N(0,1)$, $P_L(w)=\textbf{a}' \boldsymbol{\tilde{w}}$, and $\boldsymbol{\tilde{w}}'=(1,w,w^2,...,w^L)$. The matrix $A$ is positive definite by definition and $A=B'B$, where $B$ is a positive definite matrix. 

If $\textbf{c}=B\textbf{a}$, equation (\ref{cond}) becomes $\textbf{c}'\textbf{c}=1$ and $\textbf{c}'=(c_1,...,c_{L+1})$. The elements of $\textbf{c}$ can be represented using a polar coordinate transformation as $c_1=\sin{\varphi_1}$, $c_2=\cos{\varphi_1}\sin{\varphi_2},\ldots,c_{L}=\cos{\varphi_1}\ldots cos{\varphi_{L-1}}\sin{\varphi_{L}},c_{L+1}=\cos{\varphi_1}\cos{\varphi_2}\ldots\cos{\varphi_{L-1}}\cos{\varphi_{L}}$, with angles $-\pi/2\le\varphi_l\le\pi/2$, $l=1,...,L$.
The density of the latent variable in (\ref{mod2.1}) can be expressed as
\begin{equation}
    h(z|\boldsymbol{\varphi},L)=(\textbf{a}'\tilde{\textbf{z}})^2\phi(z),\label{dc}
\end{equation}
where $\textbf{a}$ can be obtained from $\textbf{c}$ as $\textbf{a}=B^{-1}\textbf{c}$, $\tilde{\textbf{z}}'=(1,z,z^2,...,z^L)$ and $\boldsymbol{\varphi}'=(\varphi_1,...,\varphi_L)$.
When $L=0$ the distribution of the latent variable reduces to the normal one.
When $L=1$, $P_L(z)=a_0+a_1z$, $a_0=\sin{\varphi_1}$, $a_1=\cos{\varphi_1}$. 
The SNP parametrization with $L=1$ includes unimodal and bimodal distributions.

\begin{figure}[H]
 \includegraphics[scale=0.5]{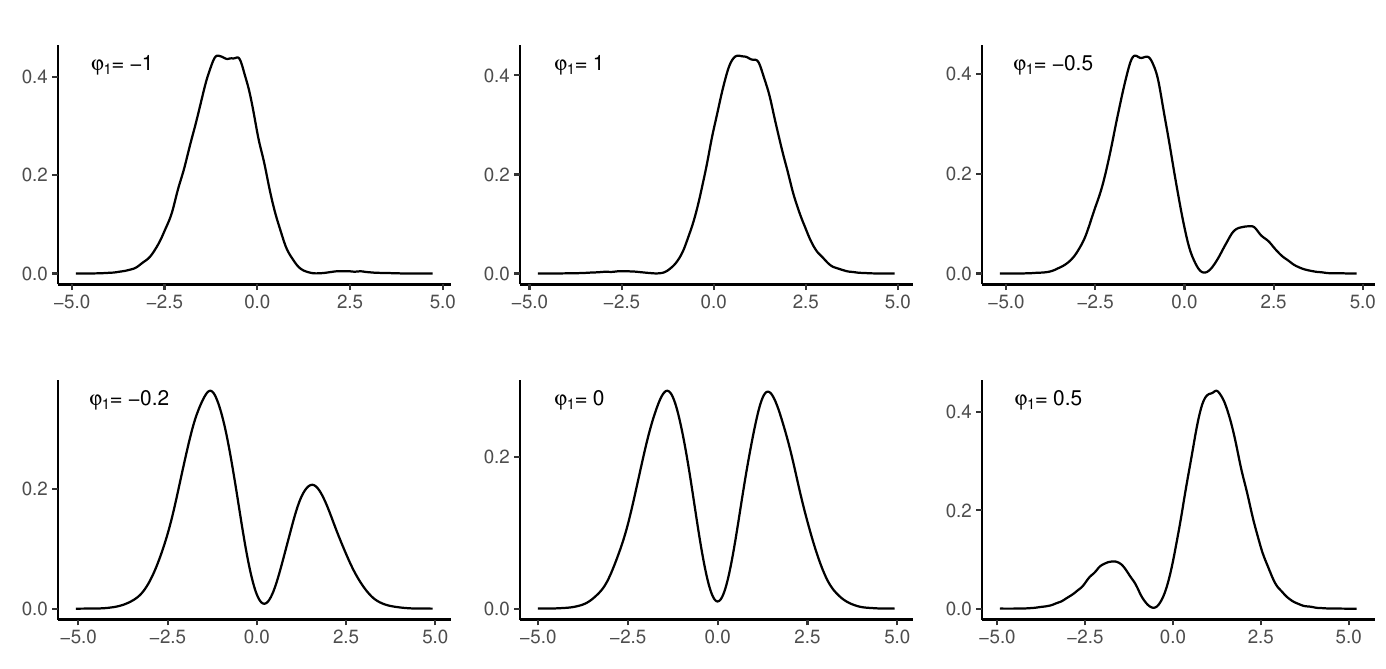}
 \caption{SNP densities of $z$ when $L=1$, for different values of the $\varphi_1$ parameter}
 \label{table:ta1}
 \end{figure}

Figure \ref{table:ta1} illustrate the SNP densities of $z$ when $L=1$, for different values of the $\varphi_1$ parameter.
When $\varphi_1$ is negative and close to -1, the distribution is slightly right-skewed, whereas when it is close to 1, it is slightly left-skewed. When the values of $\varphi_1$ are between -1 and 1, the distributions are bimodal. Even if not reported in the graph, when $\varphi_1=\pm \frac{\pi}{2}$, the SNP distribution reduces to the normal one.

When $L=2$,
$P_L(z)=a_0+a_1z+a_2z^2$, $a_0=\sin{\varphi_1}-\frac{1}{\sqrt{2}}\cos{\varphi_1}\cos{\varphi_2}$, $a_1=\cos{\varphi_1}\sin{\varphi_2}$ and $a_2=\frac{1}{\sqrt{2}}\cos{\varphi_1}\cos{\varphi_2}$. The SNP parametrization with $L=2$ is more flexible than $L=1$ and encompasses unimodal, multimodal (including up to three modes), and skewed distributions.

\begin{figure}[H]
 \includegraphics[scale=0.5]{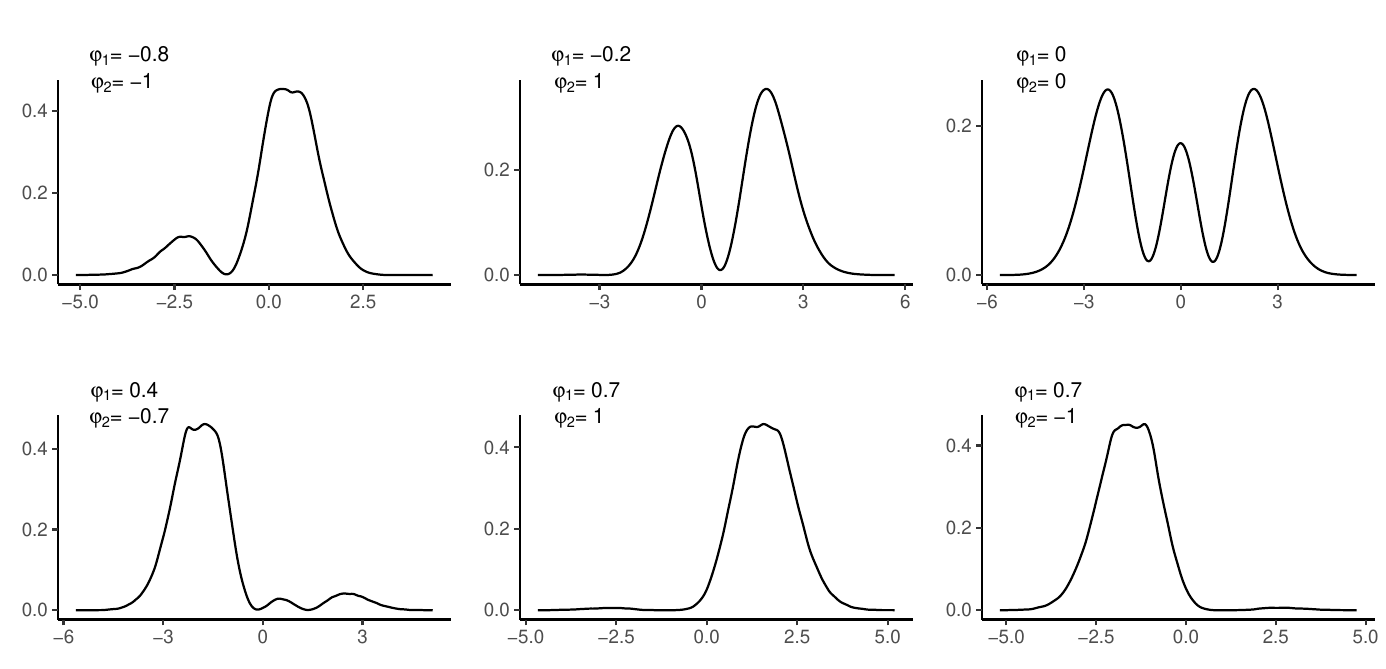}
 \caption{SNP densities of $z$ when $L=2$, for different values of the $\varphi_1$ and $\varphi_2$ parameters}
\label{table:ta2}
 \end{figure}

Figures \ref{table:ta2} displays the SNP densities of $z$ when $L=2$, for different values of the $\varphi_1$ and $\varphi_2$ parameters.

When the value of $\varphi_1$ is negative and the value of $\varphi_2$ is close to -1 or 1, the distributions are bimodal.  When both parameters are close to 0, the distributions are trimodal. When the value of $\varphi_1$ is positive and greater than 0.5, and the value of $\varphi_2$ is close to -1 or 1, the distributions are highly right-skewed and left-skewed, respectively. Even though it is not reported in Figure \ref{table:ta2}, when $\varphi_2=\pm \frac{\pi}{2}$, the SNP densities reduce to those observed when $L=1$. Moreover, when both parameters are set to $\pm \frac{\pi}{2}$, the SNP densities return to the normal case.

We indicate with $SNP_2$ the 2PL model for $L=2$, with $SNP_1$ the 2PL model for $L=1$ and with $SNP_0$ the 2PL model for $L=0$.
\subsection{Pairwise estimator for the $\boldsymbol{SNP_0}$ model}
To implement the GH test, the parameters of the $SNP_0$ model are estimated with the pairwise method.
The pairwise log-likelihood, based on the bivariate marginal densities $f({y}_{ij},y_{ik},\boldsymbol{\theta})$, $j,k=1,..p$ and $k>j$, is
\begin{equation}
\begin{split}&pl_{SNP_0}(\mathbf{y}, \boldsymbol{\theta})=\sum_{i=1}^{n}\sum_{j=1}^p\sum_{k>j}\ln f({y}_{ij},y_{ik},\boldsymbol{\theta})=\\&=\sum_{i=1}^{n}\sum_{j=1}^p\sum_{k>j}\ln \int  \bigg[\pi_{ij}(z_i)^{y_{ij}} (1-\pi_{ij}(z_i))^{1-y_{ij}}\bigg]\bigg[\pi_{ik}(z_i)^{y_{ik}} (1-\pi_{ik}(z_i))^{1-y_{ik}}\bigg]\phi(z_i)dz_i.
\end{split}\label{loklike3}\end{equation}
$pl_{SNP_0}(\mathbf{y}, \boldsymbol{\theta})$ is maximized with respect to $\boldsymbol{\theta}$, where $\boldsymbol{\theta}$ includes the item intercepts and slopes.
Under correct model specification, the maximum PL estimator $\tilde{\boldsymbol{\theta}}$ converges in probability to the true parameter vector $\boldsymbol{\theta}_0'=(\boldsymbol{\alpha}_{00}',\boldsymbol{\alpha}_{01}')$ and
\begin{equation}
    \tilde{\boldsymbol{\theta}}\xrightarrow[]{p} N(\boldsymbol{\theta}_0, A^{-1}( \boldsymbol{\theta}_0)B( \boldsymbol{\theta}_0)A^{-1}( \boldsymbol{\theta}_0)),
\end{equation}
where $A(\boldsymbol{\theta})=E_y\bigg[-\frac{\partial^2 pl_{SNP_0}(\mathbf{y},\boldsymbol{\theta})}{\partial \boldsymbol{\theta} \partial \boldsymbol{\theta}'}\bigg]$, $B=var\bigg[\frac{\partial pl_{SNP_0}(\mathbf{y},\boldsymbol{\theta})}{\partial \boldsymbol{\theta}}\bigg]$ and ${A}(\boldsymbol{\theta})\ne {B}(\boldsymbol{\theta})$ (\citealp{lindsay1988composite},
\citealp{varin2008composite}). These matrices can be estimated by their observed versions evaluated at $\tilde{\boldsymbol{\theta}}$ as
\begin{equation}
    \hat{A}({\tilde{\boldsymbol{\theta}}})=\left. \sum_{i=1}^{n}\frac{\partial^2 pl_{SNP_0}(\mathbf{y}_i,\boldsymbol{\theta})}{\partial \boldsymbol{\theta} \partial \boldsymbol{\theta}'}\right|_{\boldsymbol{\theta}=\tilde{\boldsymbol{\theta}}}\label{hess}
    \end{equation}
and \begin{equation}\hat{B}(\tilde{\boldsymbol{\theta}})=\left.\sum_{i=1}^{n}\frac{\partial pl_{SNP_0}(\boldsymbol{y}_i,\boldsymbol{\theta})}{\partial \boldsymbol{\theta}}\frac{\partial pl_{SNP_0}(\boldsymbol{y}_i,\boldsymbol{\theta})}{\partial \boldsymbol{\theta}'}\right|_{\boldsymbol{\theta}=\tilde{\boldsymbol{\theta}}}.\label{cp}\end{equation}
\subsection{Quasi-ML estimator for the $\boldsymbol{SNP_L}$ model}
The parameter vector $\boldsymbol{\theta}^{{(1)}'}=(\boldsymbol{\alpha}_0',\boldsymbol{\alpha}_1',\boldsymbol{\varphi}')=(\boldsymbol{\theta}',\boldsymbol{\varphi}')$ of the $SNP_L$ model, where $L>0$, is estimated using the quasi-ML method. The quasi-log-likelihood of the data is
\begin{equation}
\begin{split}l_{SNP_L}(\mathbf{y}, \boldsymbol{\theta}^{{(1)}})&=\sum_{i=1}^{n}\ln f(\mathbf{y}_{i},\boldsymbol{\theta}^{{(1)}})=\\&=\sum_{i=1}^{n}\ln \int \prod_{j=1}^{p} \pi_{ij}(z_i)^{y_{ij}} (1-\pi_{ij}(z_i))^{1-y_{ij}}P_L^2(z_i)\exp{\bigg(-\frac{1}{2}z_i'z_i\bigg)} dz_i.
\end{split}\label{loklike2}\end{equation}
The integral in $l_{SNP_L}(\mathbf{y}, \boldsymbol{\theta}^{{(1)}})$ is approximated by using the Gauss-Hermite quadrature, as in \citet{woods2009item}. The degree of the polynomial $L$ is fixed and is not estimated by maximum likelihood.
The quasi-log-likelihood function is maximized with respect to the unknown vector of parameter $\boldsymbol{\theta}^{{(1)}}$ as follows 
\begin{equation}(\dtt{\boldsymbol{{\alpha}}}_0',\dtt{\boldsymbol{{\alpha}}}_1',\hat{\boldsymbol{\varphi}}')=argmax_{\boldsymbol{\theta}}l_{SNP_L}(\mathbf{y}, \boldsymbol{\theta}^{{(1)}}). \label{snplf}
\end{equation}
For identifiability reasons, the item intercepts and slopes should be rescaled as described in the next section.
\subsubsection{The identifiability problem and the final estimators}
\label{ident}
The standard argument for obtaining the final form of the estimators $\hat{\boldsymbol{\alpha}}_0$ and $\hat{\boldsymbol{\alpha}}_1$ relies
on the concept of indeterminacy of the scale of the latent variable (\citealp{lord1980applications}, \citealp{van2016linking}). Indeed, in latent variable models, it is possible to change the scale of the latent variable $z$ changing the parameterization accordingly and obtain the same joint density function of the observed variables. 

After maximizing the log-likelihood in (\ref{loklike2}), the latent variable $z$ has a density $h(z|\hat{\boldsymbol{\varphi}}, L)$ with an estimated mean $\tilde{E}(Z)$ and variance $\tilde{V}(Z)$. The estimators $\dtt{\boldsymbol{\alpha}}_0$ and $\dtt{\boldsymbol{\alpha}}_1$ are on the same scale defined by $h(z|\hat{\boldsymbol{\varphi}}, L)$.
To compare the 2PL model, which assumes $z\sim N(0,1)$, with the SNP-IRT model, we rescale the mean and the variance of the latent variable of the SNP-IRT model to 0 and 1, respectively, along with adjusting the parametrization accordingly.
After the optimization process, we can express
\begin{equation}
   logit(\pi _{j}(z))=\dtt{\alpha}_{0j}+\dtt{\alpha}_{1j}z\hspace{1cm}j=1,...,p\label{transf}
\end{equation}
and \begin{equation}z=\sqrt{\tilde{V}(Z)}z_1+{\tilde{E}}({Z})\label{2}
\end{equation} where $\tilde{E}({Z})$ and $\tilde{V}({Z})$ are found given $\hat{\boldsymbol{\varphi}}$ and the SNP density of $z$ and
$z_1$ has the same distribution of $z$, but with mean 0 and variance 1.

If we substitute (\ref{2}) in (\ref{transf}) we get 
\begin{equation}
    logit(\pi _{j}(z))=\dtt{\alpha}_{0j}+\dtt{\alpha}_{1j}\sqrt{\tilde{V}(Z)} z_1+\dtt{\alpha}_{1j}\tilde{E}(Z) \label{ff} \hspace{1cm}j=1,...,p
\end{equation}
From equation (\ref{ff}) we get the form of the final estimators, which corresponds to a latent variable with mean 0 and variance 1 (\citealp{irincheeva2012generalized}):
\begin{equation}
      \hat{\alpha}_{0j}=\dtt{\alpha}_{0j}+\dtt{\alpha}_{1j}\tilde{E}(Z)\label{mu}\hspace{1cm}j=1,...,p
\end{equation}
\begin{equation}
     \hat{\alpha}_{1j}= \dtt{\alpha}_{1j}\sqrt{\tilde{V}(Z)}\label{gamma}\hspace{1cm}j=1,...,p,
\end{equation}
$\tilde{E}({Z})$ and $\tilde{V}({Z})$ can be computed analytically, given the values of $\hat{\boldsymbol{\varphi}}$, as shown in the Appendix A. Finally, $ \hat{\boldsymbol{\theta}}^{{(1)}'}=(\hat{\boldsymbol{\alpha}}^{'}_0,\hat{\boldsymbol{\alpha}}^{'}_1,\hat{\boldsymbol{\varphi}}^{'})$.

Under normal, multi-modal and asymmetric distributions of the latent variables and if the regularity conditions A2-A6 of \citeauthor{white1982maximum} (\citeyear{white1982maximum}) are satisfied, 
\begin{equation}
    \hat{\boldsymbol{\theta}}^{{(1)}}\xrightarrow[]{p} N({\boldsymbol{\theta}^{{(1)}}_{0}}, A^{-1}( {\boldsymbol{\theta}^{{(1)}}_{0}})B( {\boldsymbol{\theta}^{{(1)}}_{0}})A^{-1}( {\boldsymbol{\theta}^{{(1)}}_{0}})),
\end{equation}
 where $\boldsymbol{\theta}^{{(1)}'}_{0}=(\boldsymbol{\alpha}_{00}',\boldsymbol{\alpha}_{01}',\boldsymbol{\varphi_*}')=(\boldsymbol{\theta}_0',\boldsymbol{\varphi_*}')$. $\boldsymbol{\varphi_*}$ is the value of $\boldsymbol{\varphi}$ that minimizes the Kullback-Leibler information criterion (\citeauthor{white1982maximum}, \citeyear{white1982maximum}, \citealp{gallant1989seminonparametric}, \citealp{irincheeva2012generalized}). If the true latent variable follows an $SNP_L$ density, the vector $\boldsymbol{\varphi_*}$ coincides with the true parameter value $\boldsymbol{\varphi}_0$ and the quasi-ML method reduces to the classic full-ML method. $A(\boldsymbol{\theta})$ and $B(\boldsymbol{\theta})$ are the expected Hessian and cross-product matrices, respectively. Their observed versions can be computed with the Delta method (\citealp{cramer1946mathematical}) and are defined similarly to (\ref{hess}) and (\ref{cp}), where $pl_{SNP_0}(\mathbf{y}_i,\boldsymbol{\theta})$ is replaced by $l_{SNP_L}(\mathbf{y}_i,\boldsymbol{\theta}^{(1)})$.

\section{The Generalized Hausman Test}
In this section, we present the GH test, derived by \citet{white1982maximum}, applied here to detect non-normality of the latent variable using the SNP-IRT model.

As in the previous sections, let us denote by $\boldsymbol{\theta}$ the sub-vector of $\boldsymbol{\theta}^{{(1)}'}=(\boldsymbol{\alpha}_0',\boldsymbol{\alpha}_1',\boldsymbol{\varphi}')$ that includes the item intercepts $\boldsymbol{\alpha}_0$ and slopes $\boldsymbol{\alpha}_1$. $\boldsymbol{\theta}$ has dimension $2p\times 1$, where $p$ is the number of items. 

For the 2PL IRT model ($SNP_0$), consider the maximum PL estimator $\tilde{\boldsymbol{\theta}}_{SNP_0}$.

Consider the quasi-ML estimator $\hat{\boldsymbol{\theta}}^{{(1)}'}_{SNP_L}=({\hat{\boldsymbol{\theta}}^{'}_{SNP_L}},\hat{\boldsymbol{\varphi}}^{'})$ of a SNP-IRT model with $L>0$, where the sub-vector of parameter $\hat{\boldsymbol{\varphi}}$ has dimension $L\times 1$ and so $\hat{\boldsymbol{\theta}}_{SNP_L}^{{(1)}}$ has dimension $(2p+L)\times 1$.  
Following \citet{white1982maximum}, under normality of the latent variable
\begin{equation}
\sqrt{n}(\hat{\boldsymbol{\theta}}_{SNP_L}-\tilde{\boldsymbol{\theta}}_{SNP_0})\xrightarrow[]{d} N(0, S(\boldsymbol{\theta}_{0},{\boldsymbol{\theta}_{0}}^{{(1)}})). \label{conv}   
\end{equation} 
An estimator of $S(\boldsymbol{\theta}_{0},{\boldsymbol{\theta}_{0}}^{{(1)}})$ is given by
 \begin{equation}
 \small
\begin{split}
    &\hat{S}(\tilde{\boldsymbol{\theta}}_{SNP_0},\hat{\boldsymbol{\theta}}^{{(1)}}_{SNP_L})=\hat{A}^{\boldsymbol{\theta}\boldsymbol{\varphi}}(\hat{\boldsymbol{\theta}}^{{(1)}}_{SNP_L})^{-1}\hat{B}(\hat{\boldsymbol{\theta}}^{{(1)}}_{SNP_L})\hat{A}^{\boldsymbol{\theta}\boldsymbol{\varphi}}(\hat{\boldsymbol{\theta}}^{{(1)}}_{SNP_L})^{-1'}+\hat{A}(\tilde{\boldsymbol{\theta}}_{SNP_0})^{-1}\hat{B}(\tilde{\boldsymbol{\theta}}_{SNP_0})\hat{A}({\tilde{\boldsymbol{\theta}}_{SNP_0}})^{-1'}-\\&-\hat{A}^{\boldsymbol{\theta}\boldsymbol{\varphi}}(\hat{\boldsymbol{\theta}}^{{(1)}}_{SNP_L})^{-1}\hat{R}(\tilde{\boldsymbol{\theta}}_{SNP_0},\hat{\boldsymbol{\theta}}^{{(1)}}_{SNP_L})'\hat{A}(\tilde{\boldsymbol{\theta}}_{SNP_0})^{-1'}-\hat{A}(\tilde{\boldsymbol{\theta}}_{SNP_0})^{-1}\hat{R}(\tilde{\boldsymbol{\theta}}_{SNP_0},\hat{\boldsymbol{\theta}}^{{(1)}}_{SNP_L})  \hat{A}^{\boldsymbol{\theta}\boldsymbol{\varphi}}(\hat{\boldsymbol{\theta}}^{{(1)}}_{SNP_L})^{-1'},
    \end{split}\label{S}
\end{equation}
where the matrices 
$\hat{A}(\tilde{\boldsymbol{\theta}}_{SNP_0})$ and $\hat{B}
(\tilde{\boldsymbol{\theta}}_{SNP_0})$, defined in formulas (\ref{hess}) and (\ref{cp}), have dimension $2p\times 2p$ and are evaluated at $\tilde{\boldsymbol{\theta}}_{SNP_0}$. $\hat{A}(\hat{\boldsymbol{\theta}}^{{(1)}}_{SNP_L})$ and $\hat{B}(\hat{\boldsymbol{\theta}}^{{(1)}}_{SNP_L})$ are the observed Hessian and cross-product matrix of dimension $(2p+L)\times(2p+L)$ for the $SNP_L$ model, evaluated at $\hat{\boldsymbol{\theta}}^{{(1)}}_{SNP_L}$. The matrix $\hat{A}^{\boldsymbol{\theta}\boldsymbol{\varphi}}(\hat{\boldsymbol{\theta}}^{{(1)}}_{SNP_L})^{-1}$ is obtained by deleting the last $L$ rows from the matrix $\hat{A}(\hat{\boldsymbol{\theta}}^{{(1)}}_{SNP_L})^{-1}$ and has dimension $2p\times(2p+L)$. 
The matrix $\hat{R}(\tilde{\boldsymbol{\theta}}_{SNP_0},\hat{\boldsymbol{\theta}}^{{(1)}}_{SNP_L})$ has dimension $2p\times(2p+L)$ and can be computed as
\begin{equation}
\hat{R}(\boldsymbol{\theta}_{SNP_0},{\boldsymbol{\theta}}^{{(1)}}_{SNP_L})=\sum_{i=1}^n\frac{\partial pl_{{SNP_0}}(\boldsymbol{y}_i,\boldsymbol{\theta})}{\partial{{\boldsymbol{\theta}}}}\frac{\partial l_{{SNP_L}}(\boldsymbol{y}_i,\boldsymbol{\theta}^{{(1)}})}{\partial{\boldsymbol{\theta}}^{{(1)'}}},\hspace{0.3cm}
\label{R}
\end{equation}
where $pl_{{SNP_0}}(\boldsymbol{y}_i,\boldsymbol{\theta})$ is the pairwise log-likelihood for the individual $i$ under the model $SNP_0$ and $l_{{SNP_L}}(\boldsymbol{y}_i,\boldsymbol{\theta}^{{(1)}})$ is the log-likelihood for the individual $i$ under the model $SNP_L$. The matrix in (\ref{R}) is evaluated at $(\tilde{\boldsymbol{\theta}}_{SNP_0},\hat{\boldsymbol{\theta}}^{{(1)}}_{SNP_L})$.
We choose the maximum PL and the quasi-ML estimator for the two models to avoid that, under correct model specification, $\tilde{\boldsymbol{\theta}}_{SNP_0}$ and $\hat{\boldsymbol{\theta}}_{SNP_L}$ converge to the same covariance matrix, producing a $\hat{S}(\tilde{\boldsymbol{\theta}}_{SNP_0},\hat{\boldsymbol{\theta}}^{{(1)}}_{SNP_L})$ matrix in (\ref{S}) with all entries close to 0.

Given the theoretical result in (\ref{conv}), the GH test is given by
\begin{equation}
    GH=(\hat{\boldsymbol{\theta}}_{SNP_L}-\tilde{\boldsymbol{\theta}}_{SNP_0})'\hat{S}(\tilde{\boldsymbol{\theta}}_{SNP_0},\hat{\boldsymbol{\theta}}^{{(1)}}_{SNP_L})^{-1}(\hat{\boldsymbol{\theta}}_{SNP_L}-\tilde{\boldsymbol{\theta}}_{SNP_0}).
    \label{haus}
\end{equation}
Under normality of the latent variable, the GH test is asymptotically distributed as a $\chi^2_{2p}$, with $2p$ degrees of freedom, i.e. the number of parameters in $\boldsymbol{\theta}$. 

We consider a simpler version of (\ref{haus}), that does not involve the inversion of the matrix $\hat{S}(\tilde{\boldsymbol{\theta}}_{SNP_0},\hat{\boldsymbol{\theta}}^{{(1)}}_{SNP_L})$.
We consider the following quadratic form 
\begin{equation}
    GH_T=(\hat{\boldsymbol{\theta}}_{SNP_L}-\tilde{\boldsymbol{\theta}}_{SNP_0})'I_{2p}^{-1}(\hat{\boldsymbol{\theta}}_{SNP_L}-\tilde{\boldsymbol{\theta}}_{SNP_0}),\label{Haus2}
\end{equation}
where $I_{2p}$ can be omitted from the above formula. 

Following \cite{yuan2010two}
\begin{equation}
GH_T= \sum_{l=1}^d\lambda_l \delta_l^2, \hspace{1cm}\delta_l\sim N(0,1),\label{conv2}\end{equation}
where $d$ is the rank of $S({\boldsymbol{\theta}_{0}},{\boldsymbol{\theta}_{0}}^{{(1)}})$, and $\lambda_1,...,\lambda_d$ are its non-zero eigenvalues.

It is possible to approximate the distribution of $GH_T$ using the moment matching method (\citealp{welch1938significance},\citealp{yuan2010two})  as follows
\begin{equation}
    GH_T\sim a\chi^2_b.\label{conv3}
\end{equation}
The quantity $a$ and $b$ are defined as \begin{equation}
    a=\frac{\sum_{l=1}^{d}\lambda_l^2}{\sum_{l=1}^{d}\lambda_l}\label{a}\end{equation}
and \begin{equation}
b=\frac{(\sum_{l=1}^{d}\lambda_l)^2}{\sum_{l=1}^{d}\lambda_l^2}.\label{b}\end{equation} 
Since $S({\boldsymbol{\theta}_{0}},{\boldsymbol{\theta}_{0}}^{{(1)}})$ can be consistently estimated by $\hat{S}(\tilde{\boldsymbol{\theta}}_{SNP_0},\hat{\boldsymbol{\theta}}^{{(1)}}_{SNP_L})$ defined in (\ref{S}), $a$ and $b$ can be consistently estimated substituting $\hat{\lambda}_1,...,\hat{\lambda}_d$ in (\ref{a}) and (\ref{b}),
where $d$ is rank of $\hat{S}$ and $\hat{\lambda}_1,...,\hat{\lambda}_d$ are its non-zero eigenvalues. The approximation in (\ref{conv3}) matches the first two moments of $GH_T$ with those of $ a\chi^2_b$ (\ref{conv3}).

\section{Other goodness-of-fit tests}

In the case of normality of the latent variable, the overall goodness-of-fit of an IRT model for binary data is usually assessed through the Pearson's chi-square test and LR test (\citealp{bartholomew2011latent}). Since the $SNP_L$ and $SNP_0$ model are nested, we can consider the LR test for nested models \citep{wilks1938large}. The $SNP_L$ model, if $\varphi_1=\cdots={\varphi_L}=\pm \frac{\pi}{2}$, reduces to the $SNP_0$ model \citep{irincheeva2012generalized}. For the computation of the LR test, the $SNP_0$ model needs to be estimated using full-maximum likelihood (full-ML) instead of PL, in order to obtain a comparable value of the log-likelihood function with that of the $SNP_1$ model.

Let us denote by $\boldsymbol{\varphi}'=(\varphi_1,\cdots, {\varphi_L})$ and by $\textbf{c}'=(\pm\frac{\pi}{2},\cdots,\pm\frac{\pi}{2})$. The null and alternative hypotheses can be formulated as follows:
\begin{equation}
    H_{0}:\boldsymbol{\varphi}=\textbf{c} \hspace{0.3cm}vs\hspace{0.3cm} H_{1}:\boldsymbol{\varphi}\ne\textbf{c}.
    \label{hyp5}
    \end{equation}

    The test statistic is defined as 
\begin{equation}
    LR=2\{l_{SNP_L}(\mathbf{y}, \hat{\boldsymbol{\theta}}^{{(1)}})-l_{SNP_0}(\mathbf{y}, \hat{\boldsymbol{\theta}})\},
    \label{lr}
    \end{equation}
    where $l_{SNP_L}(\mathbf{y}, \hat{\boldsymbol{\theta}}^{{(1)}})$ and $l_{SNP_0}(\mathbf{y}, \hat{\boldsymbol{\theta}})$ are the quasi-log-likelihood and full-log-likelihood functions of the $SNP_L$ and $SNP_0$ models, respectively, evaluated at their maximum values. 
   Under $H_0$, the LR test is asymptotically distributed as a $\chi^2_L$.

However, the LR test is affected by the problem of sparseness. Indeed, for a fixed sample size, the number of empty cells in a frequency table increases with the number of binary items. In this case, the distribution of the LR test statistic is not well approximated by the chi-square distribution. To overcome this problem,
\citet{maydeu2005limited} propose a family of test statistics $M_r$, based on the residuals up to order $r$. The most popular statistic is $M_2$, that uses the univariate and bivariate marginal information. As data sparseness increases, the empirical Type I error rates of the $M_2$ test remain accurate (\citeauthor{maydeu2005limited}, \citeyear{maydeu2005limited}, \citeyear{maydeu2006limited}). 
As before, let's consider $p$ items and a sample size $n$. Under the null hypothesis we test that the $SNP_0$
model holds. The hypotheses $H_{0}$ and $H_{1}$ can be formalized as follows:
\begin{equation}
    H_{0}:\boldsymbol{\pi}=\boldsymbol{\pi}(\boldsymbol{\theta})\hspace{0.5cm}vs\hspace{0.5cm}H_{1}:
    \boldsymbol{\pi}\ne\boldsymbol{\pi}(\boldsymbol{\theta}),
    \label{hyp4}
    \end{equation}
where $\boldsymbol{\theta}$, as usual, includes the item intercepts and slopes and $\boldsymbol{\pi}(\boldsymbol{\theta})$ indicates the response patterns probabilities. 

The statistic $M_2$ is (\citeauthor{maydeu2005limited}, \citeyear{maydeu2005limited}):
\begin{equation}
    M_2=n\hat{\textbf{e}}_2'\hat{\boldsymbol{U}}_2\hat{\textbf{e}}_2.\label{M2}
\end{equation}
The vector $\hat{\textbf{e}}$ includes the univariate and bivariate residuals while the matrix $ \hat{\boldsymbol{U}}_2$ depends on a transformation matrix and on the Jacobian matrix of the cell probabilities with respect to
the items intercept and slope parameter
(more details can be found in \citeauthor{maydeu2005limited}, \citeyear{maydeu2005limited}, \citeyear{maydeu2006limited}). Under $H_{0}$, the statistic $M_2$ is asymptotically distributed as a $\chi^{2}_m$, with degrees of freedom $m=\frac{p(p+1)}{2}-2p$, that is the number of univariate and bivariate residuals minus the number of estimated parameters of the $SNP_0$ model. 
In the simulations, we evaluate the performance of the LR and $M_2$ tests under
non-normality of the latent variable.

\section{Model selection criteria}
The Akaike information criterion (AIC), the Bayesian information criterion (BIC) and the Hannan–Quinn criterion (HQ) can be used to choose the degree of the polynomial $L$ of the SNP-IRT model (\citealp{davidian1993nonlinear}, \citealp{woods2009item}, \citealp{irincheeva2012generalized}, \citealp{monroe2014multidimensional}).

The AIC is (\citealp{akaike1974new}):
\begin{equation}
    AIC=-2l (\mathbf{y}, \hat{\boldsymbol{\theta}}^{(1)})+2k,
\end{equation}
where $l (\mathbf{y}, \hat{\boldsymbol{\theta}}^{(1)})$ is the quasi-log-likelihood function of the $SNP_L$ model evaluated at the maximum value and $k$ is the number of parameter of the model.

The BIC is (\citealp{schwarz1978estimating}): 
\begin{equation}
    BIC=-2l (\mathbf{y}, \hat{\boldsymbol{\theta}}^{(1)})+k\ln{n},
\end{equation}
where $n$ is the sample size.

The HQ is (\citealp{hannan1987rational}):
\begin{equation}
   HQ=-2l (\mathbf{y}, \hat{\boldsymbol{\theta}}^{(1)})+2k\ln{\ln{n}}
\end{equation}
Usually $L=1$ or $L=2$ are enough to detect a departure from normality and to approximate different shapes of latent variable distributions. Selecting higher order of the polynomial could result in overfitting (\citealp{irincheeva2011generalized}).

When $L=0$, $\hat{\boldsymbol{\theta}}^{(1)}=\hat{\boldsymbol{\theta}}$ in the above formulas and, as for the LR test, the $SNP_0$ model is estimated using full-ML.
\section{Simulation study}
\label{studyGH}
\subsection{Design}
In this section, we study the performance of the $GH_T$ test to assess non-normality of the latent variable distribution and we compare its performance with the $M_2$ and $LR$ tests. Moreover, for all simulation scenarios, the BIC, AIC and HQ criteria have been computed.

To evaluate the performance of the $GH_T$, $M_2$ and $LR$ tests, we have considered five scenarios (SC), corresponding to five different distribution assumptions for the latent variable $z$ in the data generating models.
The general model is 
\begin{equation}
\begin{split}
&logit(\pi_{ij})=\alpha_{0j}+\alpha_{1j}z_i\hspace{1cm} i=1,...,n\hspace{1cm} j=1,2,...,p\\ &z \sim f(z)
\end{split}
\label{no}
\end{equation}
Item intercepts have been randomly chosen from the interval [-0.8; 1.12] while the item slopes from the interval [0.5; 1.5].
To study the Type I error rates of the  $GH_T$, $M_2$ and LR tests we have considered the following scenario:
\begin{enumerate}[label=\textbf{A},  series=A]\item\label{A11}
$z\sim N(0,1)$
\end{enumerate}
To study the power of the  $GH_T$, $M_2$ and LR tests we have considered the following cases of two bimodal distributions and two skewed distributions:
\begin{enumerate}[label=\textbf{B},  series=B] \item\label{B11} $z \sim0.7N(-1,0.7)+0.3N(1,0.8),$

 where $z$ has an overall mean equal to -0.40 and variance equal to 1.38.
 \end{enumerate}
\begin{enumerate}[label=\textbf{C},  series=C] \item \label{C11} $z \sim0.1N(-2,0.25)+0.9N(2,1),$

 where $z$ has an overall mean equal to 1.6 and variance equal to 2.37.
\end{enumerate}
\begin{enumerate}[label=\textbf{D},  series=D] \item \label{D11} $Z\sim SN(\mu=-2.5,\sigma=2,\lambda=5)$, 

 where $z$ has mean -0.93 and variance 1.55.

\end{enumerate}
\begin{enumerate}[label=\textbf{E},  series=E] \item \label{E11}  $Z\sim SN(\mu=-2.5,\sigma=2,\lambda=10)$, 

 where $z$ has mean -0.91 and variance 1.47.

\end{enumerate}

The mixture of normals in scenario \ref{C11} wider departs from the normal distribution than the one in scenario \ref{B11}. Similarly, the skew-normal distribution in scenario \ref{E11} exhibits a greater departure from the normal distribution than the one in \ref{D11}.

Two versions of the $GH_T$ test have been considered in the simulations. The first version, denoted as $GH_{T1}$, is based on the $SNP_0$ and $SNP_1$ models. The choice of the $SNP_1$ model has been motivated by the fact that it can approximate bimodal distributions well, such as the ones in scenarios \ref{B11} and \ref{C11}.

The optimization of the $SNP_1$ model has been achieved in R with direct maximization via
the function “nlminb”, that uses the analytically computed gradient and Hessian matrix. For the $SNP_1$ model, the initial values of the parameters $\boldsymbol{\alpha}_0$ and $\boldsymbol{\alpha}_1$ used in the optimization process are the full-ML parameter estimates obtained with the $SNP_0$ model. In each data replication, for the $\varphi_1$ parameter, we have sampled 10 initial values from a sequence of values, equally spaced by 0.1 in the interval $[-\frac{\pi}{2};\frac{\pi}{2}]$, i.e. the domain of $\varphi_1$. Among the estimated $SNP_1$ models in each data replication, we have selected the one that corresponds to the maximum value of the quasi-log-likelihood function.
The performance of the $GH_{T1}$ test has been evaluated under all scenarios.

In scenarios \ref{D11} and \ref{E11}, a second version of $GH_T$ called $GH_{T2}$ has been considered in addition to $GH_{T1}$. $GH_{T2}$ is based on the $SNP_0$ model and the $SNP_2$ model.  This choice is motivated by the fact that the SNP model provides a representation that captures the highly skewed case particularly well when $L=2$, as showed in Figure \ref{table:ta2}. As for the $SNP_1$ model, also the optimization of the $SNP_2$ model has been conducted in R using the direct maximization function "nlminb". However, the $SNP_2$ model is highly sensitive to the initial values chosen. To mitigate this, the true parameter values have been used as initial values for the $\boldsymbol{\alpha}_0$ and $\boldsymbol{\alpha}_1$ parameters. The initial values for the $\varphi_1$ and $\varphi_2$ parameters have been set to 0.7 and 1, respectively, which correspond to a skew-normal case.
To compute the $GH_{T1}$ and $GH_{T2}$ tests, the $SNP_0$ model has been estimated using the maximum PL method 
through the R function "optim".  The Hessian, cross-product matrices and the matrix in formula (\ref{R}) involved in the $GH_{T1}$ and $GH_{T2}$ tests have been computed numerically with the “NumDeriv” R package. 

To compute the values of the information criteria and the $LR$ test, the $SNP_0$ model has been estimated using the full-ML method. This estimation has been achieved in R via the function "optim". The $M_2$ test in each data replication has been computed using the "M2" function of the R package "mirt".

The Type I error rates and power of the $GH_{T1},GH_{T2}$, $M_2$ and LR tests have been computed as follows: $\hat{p}=\sum_{l=1}^{N_v}\frac{I(T_{l} \ge c)}{N_v}$, where $N_v$ represents the number of valid statistics out of the total replications. Here, $I$ denotes an indicator function, and $T_{l}$ is the value of the test statistic evaluated in the $l$-th replication. The critical value $c$ corresponds to the theoretical asymptotic value, specifically the $(1-\alpha)$th percentile of the $\hat{a}\chi^2_{\hat{b}}$ distribution for the $GH_T$ test. The values $\hat{a}$ and $\hat{b}$ are computed as in (\ref{a}) and (\ref{b}). For the $M_2$ test, the critical value is associated with the $\chi^2_{{m}}$ distribution, where $m$ is equal to $\frac{p(p+1)}{2}-2p$.

Considering that the $SNP_1$ model has one more parameter than the $SNP_0$ model, the LR test uses $c$ as the theoretical asymptotic critical value corresponding to the $(1-\alpha)$th percentile of the $\chi^2_{{1}}$ distribution. This specific test is referred to as $LR_1$. For the comparison of $SNP_0$ with $SNP_2$, the LR test with 2 degrees of freedom is used, denoted as $LR_2$.

To establish the confidence interval (CI) of each rate $\hat{p}$, we calculate it as $\hat{p}\pm z_{(1-\frac{\alpha}{2})}\sqrt{\frac{\alpha(1-\alpha)}{N_v}}$.

Next, the percentages of times the AIC, BIC, and HQ criteria select the $SNP_1$ over $SNP_0$ have been computed as $\hat{P}=\sum_{l=1}^{N_v}\frac{I(IC_{SNP_1}<IC_{SNP_0})}{N_v}100$, where IC indicates the AIC, BIC, and HQ criteria. Similarly, we computed percentages for the selection of the $SNP_2$ model over the $SNP_0$ model using information criteria.

For scenarios \ref{A11}, \ref{B11} and \ref{C11}, we have considered the following simulation conditions: number of items $(p=10,20)$ $\times$ sample size ($n=500, 1000, 5000$)$\times$ test statistics ($GH_{T1}$, $M_2$, $LR_1$)$\times$ information criterion (AIC, BIC, HQ). For scenarios \ref{D11} and \ref{E11}, the simulation conditions are: number of items $(p=10,20)$ $\times$ sample size ($n=500, 1000$)$\times$ test statistics ($GH_{T1}$,$GH_{T2}$, $M_2$, $LR_1$, $LR_2$)$\times$ information criterion (AIC, BIC, HQ). In all the simulation scenarios, $R=500$ replications and two nominal levels of $\alpha$ have been considered, that is $\alpha=0.05,0.01$. 

Preliminary results on the performance of the $GH_{T1}$ test under scenario \ref{A11} and \ref{C11} have been obtained in \cite{guas2023}. The test showed good performance both in terms of Type I error rates and power, especially for large sample sizes.

Some additional results, that include the bias for the parameter estimates of the $SNP_0$, $SNP_1$ and $SNP_2$ model, have been reported in the Appendix B. If the mean of the true latent variable is different from 0 and the variance different from 1, to compute the parameter bias, the true $\boldsymbol{\alpha}_{00}$ and $\boldsymbol{\alpha}_{01}$ parameters have been rescaled accordingly to formula (\ref{mu}) and (\ref{gamma}), replacing $\dtt{\alpha}_{0j}$ with the true intercept, $\dtt{\alpha}_{1j}$ with the true slope, $\tilde{E}(Z)$ and $\tilde{V}(Z)$ with the mean and the variance of the true latent variable. In this way, the parameter bias is due only to the misspecification of the shape of the latent variable, and not to the misspecification of the moments. A similar procedure to compute parameter bias has been adopted by \citet{irincheeva2011generalized}.

\subsection{Results}
Table \ref{tab2.1} reports the Type I error rates of the $GH_{T1}$, $M_2$ and $LR_1$ tests
for scenario \ref{A11}.
\begin{table}[H]
\caption{Scenario A: Type I error rates of the $GH_{T1}$, $M_2$ and $LR_1$ tests, $p=10,20$, $n=500,1000,5000$ }  
\centering  
\relsize{-0.5}
\begin{tabular}{c ccccccc ccc}  
\hline\hline
&&\multicolumn{3}{c|}{$\alpha=0.05$}& \multicolumn{3}{|c}{$\alpha=0.01$} \\\cline{3-8}
$p$ & $n$&$GH_{T1}$& $M_2$&$LR_1$& $GH_{T1}$& $M_2$&$LR_1$\\ [0.5ex]   \hline

10&500  & \textbf{0.018}& 0.066
&\textbf{0}&0.002& 0.012
&0

\\
&1000  &0.044&0.046
&\textbf{0.002}
&0.006&0.01&0\\
&5000  &0.056& 0.04
&0.042
&0.02&0.002
&0

\\\\
20&500&\textbf{0.026}& 0.036&\textbf{0.006}& 0.008&0&0

\\
&1000& 0.042& 0.054
& 0.054
&0.008&0.006
&0

\\
&5000&0.042 &0.05
&\textbf{ 0.264
}&0.01&0.016
&\textbf{0.094}

\\

\hline\hline
\end{tabular}

\relsize{-1}
\footnotetext[1]{} Note 1: Values in boldface indicate that the nominal level $\alpha$ is not included in their confidence interval\\
\label{tab2.1}
\end{table}

Overall, the $GH_{T1}$ test has good performance in terms of Type I error rates under most conditions. Indeed, $GH_{T1}$ is more conservative than expected only for $\alpha=0.05$, 10 and 20 items and small sample size. $M_2$ has good performance in terms of Type I error rates for all values of $\alpha$, number of items and sample sizes considered. Among the three tests considered, the $LR_1$ test has the worst performance. For $\alpha=0.05$, it rejects less than it should with 10 items, $n=500,1000$ and 20 items, $n=500$. Moreover, for all level of $\alpha$ considered, it has seriously inflated Type I error rates with 20 items and $n=5000$.

Table \ref{tab2.1.1} shows the percentages of times AIC, BIC and HQ select $SNP_0$ instead of $SNP_1$ for scenario \ref{A11}.
\begin{table}[H]
\caption{Scenario A: percentages of times AIC, BIC and HQ select $SNP_0$ instead of $SNP_1$, $p=10,20$, $n=500,1000,5000$}  
\centering  
\relsize{-0.5}
\begin{tabular}{c c ccccccc}  
\hline\hline
$p$ & $n$& AIC &BIC &HQ \\ [0.5ex]   \hline

10&500  &  97\%&100\%&99.8\%

\\ 
&1000&  93.2
\%&100\%&98.8\%\\
&5000&  90.6

\%&100\%& 96.6
\%
\\\\
20&500 &87.8\%&99.8\%&99.2\%

\\
&1000&  78.4\%&100\%&94.6
\%
\\
&5000&  54.4\%&97
\%&78.4\%
\\
\hline\hline
\end{tabular}

\label{tab2.1.1}
\end{table}

Among the three information criteria considered, the AIC has the worst performance. This is evident especially with 20 items and all sample sizes, where it selects the $SNP_0$ model more or less from 54\% to 88\% of times. The BIC has the best performance and it selects the $SNP_0$ model almost in the totality of cases under all conditions. The performance of the HQ to select the $SNP_0$ model is in between the performance of the AIC and BIC criteria.

Table \ref{tab2.2} presents the empirical power of the $GH_{T1}$, $M_2$ and $LR_1$ tests for scenarios \ref{B11} and \ref{C11}.
\begin{table}[H]
\caption{Scenarios B and C: empirical power of the of the $GH_{T1}$, $M_2$ and $LR_1$ tests, $p=10,20$, $n=500,1000,5000$}  
\centering  
\relsize{-0.5}
\begin{tabular}{c ccccc cccccc}  
\hline\hline
&&&\multicolumn{3}{c|}{$\alpha=0.05$}& \multicolumn{3}{|c}{$\alpha=0.01$}  \\\cline{4-9}
SC&$p$ & $n$& $GH_{T1}$& $M_2$&$LR_1$&$GH_{T1}$& $M_2$&$LR_1$\\\hline

\ref{B11}&10&500&0.612&0.03&0.572
& 0.536&0.004
& 0.462
 \\
&&1000 &0.868
 
&0.028
& 0.792&0.796&0.006

& 0.558
\\&&5000 &1& 0.052&0.936&1&0.014&0.93

\\\\
&20&500  & 0.886
&0.022
&0.596&0.76& 0.006&0.428

\\
&&1000  &0.984&0.036&0.63&0.96& 0.002&0.60\\
&&5000  & 0.996&0.034&0.678
&0.996&0.004& 0.666

\\\\

\ref{C11}&10&500 &0.924&0.01&0.912&0.852&0.002&0.8
\\
&&1000&0.998&0.006&0.968&0.994&0&0.96\\
&&5000&1&0.028&0.972&1& 0.004&0.972
\\\\
&20&500  &0.988 &0&0.782& 0.95&0&0.72

\\
&&1000&0.996&0& 0.822&0.994&0&0.756\\
&&5000&0.998& 0.004&0.922
& 0.998&0&0.892

\\

\hline\hline
\end{tabular}\label{tab2.2}

\end{table}

Among the three tests and under all levels of $\alpha$ considered, $GH_{T1}$ has the highest power when the true latent variable is generated from a mixture of normals, that is under scenarios \ref{B11} and \ref{C11}. In general, the power of the $GH_{T1}$ and $LR_1$ tests increases as the sample size and the number of items increase and as the true latent variable distribution largely departs from the normal one. For what concerns the $M_2$ test, it has very low or no power to detect non-normality of the latent variable distribution under both scenarios considered.

Table \ref{tab2.22} shows the percentages of times AIC, BIC and HQ select $SNP_1$ instead of $SNP_0$ for scenarios \ref{B11} and \ref{C11}.
\begin{table}[H]
\caption{Scenarios B and C: percentages of times AIC, BIC and HQ select $SNP_1$ instead of $SNP_0$, $10,20$, $n=500,1000,5000$}  
\centering  
\relsize{-0.5}
\begin{tabular}{c c ccccccc}  
\hline\hline
SC&$p$ & $n$& AIC &BIC &HQ\\ [0.5ex]   \hline

\ref{B11}&10&500  &75.6
\%&47.6\%& 58.8
\%

\\
&&1000&85.2\%&55.4
\%&78.8
\%\\
&&5000& 94\%&93\%&93.4\%\\\\
&20&500 &66.2
\%&47\%&60.2
\%

\\
&&1000&65.2\%&60\%&63\%\\
&&5000& 70\%&65.4
\% &67.6
\%\\\\

\ref{C11}&10&500 &95.6\%&80.6\%&91.8\%

\\
&&1000&  97\%&95.6\%&96.8\%
\\
&&5000&  97.6\%&97.2\%&97.2\%
\\\\
 &20&500 &85.8\%&73\%&79.4\%
\\

&&1000&87.6\%&75.4\%&82\%\\
&&5000&95\%&87.2\%&92
\% \\

\hline\hline
\end{tabular}\label{tab2.22}

\end{table}

Under all scenarios, AIC has the best performance when $n=500,1000$ and it has the highest percentages of selections as the $SNP_1$ model. Under scenarios \ref{B11} and \ref{C11}, AIC has the best performance also for $n=5000$. In the majority of cases, BIC has the worst performance when $n=500,1000$. This is evident especially under scenario \ref{B11} and $n=500$, where it selects the $SNP_1$ models only around 47\% of times. The performance of HQ to select the $SNP_1$ model under all scenarios for small sample size is in between the performance of AIC and BIC criteria.

Table \ref{tab2.23} reports the empirical power of the $GH_{T1}$, $LR_1$, $GH_{T2}$, $LR_2$ and $M_2$ tests for scenarios \ref{D11} and \ref{E11}.
\begin{table}[H]
\caption{Scenarios D and E: empirical power of the $GH_{T1}$, $LR_1$, $GH_{T2}$, $LR_2$ and $M_2$ tests, $p=10,20$, $n=500,1000$ }  
\centering  
\relsize{-0.5}
\begin{tabular}{c cccccccccc cc}  
\hline\hline
&&&\multicolumn{5}{c|}{$\alpha=0.05$}& \multicolumn{5}{|c}{$\alpha=0.01$} \\\cline{4-13}
SC&$p$ & $n$& $GH_{T1}$&$LR_1$&$GH_{T2}$&$LR_2$&$M_2$&$GH_{T1}$ &$LR_1$&$GH_{T2}$ &$LR_2$&$M_2$\\ [0.5ex]   \hline

\ref{D11}&10&500  &0.554&0.374&0.492&0.7&0.024&0.416&0.27&0.338&0.462&0.004\\
&&1000 &0.764&0.588&0.722&0.956&0.028&0.636&0.264&0.608&0.85& 0.008\\\\
&20&500  &0.786&0.358&0.89&0.972&0.026&0.576&0.196&0.762&0.876&0.008\\
&&1000 &0.962&0.38&0.992&1&0.02&0.924&0.348&0.972&1&0.002\\\\
\ref{E11}&10&500  &0.588&0.43&0.562&0.818&0.028& 0.458&0.334&0.404&0.62&0.01\\
&&1000 &0.894&0.736&0.814&0.98&0.036&0.834&0.448&0.688&0.92&0.014\\\\
&20&500  &0.874&0.45&0.916&0.99&0.022&0.71&0.282& 0.836&0.946&0.006\\
&&1000 &0.986&0.474&1&1&0.026&0.948&0.458&0.998&1&0.026\\
\hline\hline
\end{tabular}

\relsize{-1}

\label{tab2.23}
\end{table}
Under scenarios \ref{D11} and \ref{E11}, with 10 items, $GH_{T1}$ exhibits a slightly higher power compared to $GH_{T2}$. As reported in the Appendix B, the $SNP_1$ and $SNP_2$ models result in a similar reduction in parameter bias compared to $SNP_0$. However, $SNP_2$ better approximates the shape of the true latent variable distribution (see Appendix B) and with 20 items, the power of the $GH_{T2}$ test is slightly higher compared to $GH_{T1}$. In general, both tests achieve a high power with large sample sizes and as the skew-normal becomes more extreme.
$LR_1$ has low power under all conditions, while $LR_2$ exhibits the highest power among all scenarios. Although increasing the degree of the polynomial can enhance the $LR$ and $GH_T$ tests performance, the $SNP_2$ method requires accurate initial values for the parameters to yield good results, making it challenging to use in practice.
As observed in scenarios \ref{B11} and \ref{C11}, the $M_2$ test has very low or no power in detecting misspecifications of the latent variable distribution.

Table \ref{tab2.24} shows the percentages of times AIC, BIC and HQ select $SNP_1$ over $SNP_0$ and $SNP_2$ over $SNP_0$ for scenarios \ref{D11} and \ref{E11}.
\begin{table}[H]
\caption{Scenarios D and E: percentages of times AIC, BIC and HQ select $SNP_1$ over $SNP_0$ and $SNP_2$ over $SNP_0$, $p=10,20$, $n=500,1000$}  
\centering  
\relsize{-0.5}
\begin{tabular}{c c cccccccc}  
\hline\hline
&&&\multicolumn{3}{c|}{$SNP_1$ over $SNP_0$}& \multicolumn{3}{|c}{$SNP_2$ over $SNP_0$} \\\cline{4-9}
SC&$p$ & $n$& AIC &BIC &HQ& AIC &BIC &HQ& \\ [0.5ex]   \hline

\ref{D11}&10&500  &58.4\%&28.4\%&39\%&87\%&26.2\%&60.4\%

\\ 
&&1000  &70.4\%&24.4\%&58.4\%&99\%&61.8\%&90\%\\\\
&20&500  &44.8\%&22\%&36.6\%&98.8\%&75.4\%&94\%\\
&&1000  &41.4\%&34.6\%&38\%&100\%&98.2\%&100\%\\\\
\ref{E11}&10&500  &64.6\%&34.4\%&44.2\%&91.6\%&42.2\%&72.6\%

\\ 
&&1000  &81.2\%&44\%&73.2\%& 99\%&76\%&95\%\\\\
&20&500  &51.2\%&31\%&45.2\%&99.6\%&84.4\%&97.8\%\\
&&1000  &49.6\%&45.6\%&47.4\%&100\%&99.6\%&100\%\\
\hline\hline
\end{tabular}

\label{tab2.24}
\end{table}
 None of the criteria have good performance in selecting $SNP_1$ over $SNP_0$ for both scenarios and considered sample sizes. 
However, since the true latent variables are skewed and $SNP_2$ better approximates this type of distributions, the performance of all the information criteria improves when selecting between $SNP_0$ and $SNP_2$, with AIC showing the best performance and BIC performing the worst. Overall, from the results in Table \ref{tab2.1.1}, Table \ref{tab2.22} and Table \ref{tab2.24}, it has not been possible to identify the criterion that has the best performance under normality and non-normality of the latent variable.

\section{Real data application: American students exposure to school and neighbourhood violence}
\label{dataNLSF}
 This research is based on data from the National Longitudinal Survey of Freshmen (NLSF), a project designed by Douglas S. Massey and Camille Z. Charles and funded by the Mellon Foundation and the Atlantic Philanthropies (available at \\http://oprdata.princeton.edu/archive/restricted). The aim of the NLSF is to collect data to explain the minority underachievement in higher education. Data have been collected from 1999 to 2003, in four waves, to capture emergent psychological processes, measuring the degree of social integration and intellectual engagement. The survey included equal-sized samples of white, black, Asian, and Latino freshmen entering selective colleges and universities. We have analyzed only a part of the questionnaire that refers to the year 1999. In particular we have selected 9 binary items that measure violence in the neighbourhood. The items description is reported in Table \ref{items}. 
\begin{table}[H]
\caption{NLSF data: item description}
\centering\begin{tabular}{cc}  
\hline \hline
Item&Question\\ \hline
  1 &  \multicolumn{1}{l}{In your neighborhood, before you were ten}\\
   &\multicolumn{1}{l}{do you remember seeing homeless people on the street?}\\
2& \multicolumn{1}{l}{Prostitutes on street?}\\
3& \multicolumn{1}{l}{Gang members hanging out on the street?}\\
4& \multicolumn{1}{l}{Drug paraphernalia on the street?}\\
 5 & \multicolumn{1}{l}{ People selling illegal drugs in public?}\\
6& \multicolumn{1}{l}{People using illegal drugs in public?}\\
7& \multicolumn{1}{l}{People drinking or drunk in public?}\\
8& \multicolumn{1}{l}{Physical violence in public?}\\
9& \multicolumn{1}{l}{Hearing the sound of gunshots?}\\\hline \hline
\end{tabular}
\label{items}
\end{table}
The original sample of observations is composed by 3924 observations. Possible responses are “no”, “yes”, “don't know” and “refused”. Individuals that have responded “don't know” or “refused” have been excluded from the analysis, while responses “no” have been coded as 0 and “yes” as 1. The data set that has been analyzed is finally composed by 3891 observations. A sub-sample of 400 observations and a larger number of items of the questionnaire that refers to the year 1999 has been analyzed also by \citet{cagnone2012factor}, that fitted a latent trait models with two factors distributed as a mixture of normals. They found that the item reported in Table \ref{items} are highly loaded only on one factor, distributed as a mixture of normals.

Some preliminary descriptive analysis on this dataset have been performed with the “ltm” R package.  Even if not reported in the tables, all items are statistically associated and in all the items the proportion of “0” responses is higher than “1”. 

 The
first step of the analysis has been to fit the $SNP_0$ and $SNP_1$ models to the data.
Table \ref{tab3.6} reports the full-ML and quasi-ML parameter estimates of the $SNP_0$ and $SNP_1$ models, respectively, and related standard errors, based on the sandwich covariance matrix.
\begin{table}[H]
\caption{NLSF data: $SNP_0$ and $SNP_1$ parameter estimates and related standard errors }
\centering\begin{tabular}{cccc}  
\hline\hline
&\multicolumn{2}{c}{Estimates}\\\cline{2-3}
Parameter & $SNP_0$ 
&$SNP_1$\\\hline
$\alpha_{01}$&-1.85 (0.07)&
-2.61 (0.13)\\
$\alpha_{02}$&-5.21 (0.22)&
-6.56 (0.34)\\
$\alpha_{03}$&-3.43 (0.14)&
-4.88 (0.25)\\
$\alpha_{04}$&-5.05 (0.27)&
-7.49 (0.52)\\
$\alpha_{05}$&-6.85 (0.48)&
-9.43 (0.68)\\
$\alpha_{06}$&-5.75 (0.31)&
-7.76 (0.46)\\
$\alpha_{07}$&-1.61 (0.09)&
-2.78 (0.17)\\
$\alpha_{08}$&-1.99 (0.09)&
-2.99 (0.16)\\
$\alpha_{09}$&-2.76 (0.10)&
-3.68 (0.18)\\
$\alpha_{11}$&1.94 (0.09)&
2.93 (0.18)\\
$\alpha_{12}$&2.40 (0.15)&
4.09 (0.28)\\
$\alpha_{13}$&2.84 (0.15)&
4.64 (0.26)\\
$\alpha_{14}$&3.77 (0.23)&
6.71 (0.47)\\
$\alpha_{15}$&4.56 (0.37)&
7.75 (0.57)\\
$\alpha_{16}$&3.38 (0.22)&
5.83 (0.38)\\
$\alpha_{17}$&2.77 (0.15)& 
4.16 (0.26)\\
$\alpha_{18}$&2.32 (0.11)&
3.58 (0.21)\\
$\alpha_{19}$&2.00 (0.10)&
3.19 (0.19)\\
$\varphi_{1}$& -&0.23(0.04)\\
\hline \hline
\end{tabular}\label{tab3.6}
\end{table}

Despite both methods being on the same scale, we notice dissimilar parameter estimates. \cite{irincheeva2011generalized} highlights that when the $SNP_0$ and $SNP_L$ methods are employed on real binary data with true latent variables that may deviate from a normal distribution, they yield significantly different parameter estimates. The $SNP_0$ model might not be sufficient to effectively capture the data, resulting in notable bias in both the item intercepts and slopes (\citealp{ma2010explicit}).

To choose the best model for the data, first we have evaluated the fit of the $SNP_0$ model. 

Since the data are sparse and 174 observed response patterns out of the total 231 observed response patterns have expected frequencies less than 5, residuals calculated from marginal frequencies have been inspected. We have considered the rule of thumb that residuals greater than 4 are indicators of bad fit of the correspondent pair or triplets of items  (\citealp{bartholomew2011latent}). Even if not reported in the tables, the $SNP_0$ model does not have a good fit for some pairs and triplets of items. To evaluate if the $SNP_1$ model has a better fit than the $SNP_0$ model, information criteria have been computed.

Table \ref{tab3.7} reports the values of the AIC, BIC and HQ criteria for the $SNP_0$ and $SNP_1$ models.
\begin{table}[H]
\caption{NLSF data: Information criteria for the $SNP_0$ and $SNP_1$ models}
\centering\begin{tabular}{cccc}  
\hline\hline
          & AIC   &   BIC     &  HQ\\\hline
$SNP_0$ &21309.32 &21422.12 &21349.36\\
$SNP_1$& 21303.67 &21422.73&21345.93\\
\hline\hline
\end{tabular}\label{tab3.7}
\end{table}

The information criteria give conflicting results. Indeed AIC and HQ select the $SNP_1$ model, while BIC the $SNP_0$ model. 

In coherence with the simulation study, the $M_2$, $LR_1$ and $GH_{T1}$ tests have been computed. 

Table \ref{tab3.5} reports the value of the $M_2$, $LR_1$ and $GH_{T1}$ statistics, the degrees of freedom (d.o.f) and the associated $p$-values.

\begin{table}[H]
\caption{NLSF data: $M_2$, $LR_1$ and $GH_{T1}$ test statistics and associated degrees of freedom and $p$-values}
\centering\begin{tabular}{cccc}  
\hline\hline
Test & Value & d.o.f &$p$-value\\\hline
$M_2$&182.33            &   27    &  0\\
 $LR_1$ & 7.66&   1&  0.006\\
    $GH_{T1}$    &     222.92     & 2.65  & 0\\   
 \hline\hline
\end{tabular}\label{tab3.5}
\end{table}
According to the $M_2$ test, the $SNP_0$ model does not have a good fit to the data. However, the $M_2$ test does not reveal the source of misfit. It could be the non-normality of the latent variable even if, as showed in the simulation study, the $M_2$ test has a very low power to detect this type of model misspecification, or other types of model misspecification. 
According to the $LR_1$ and $GH_{T1}$ tests, the null
hypothesis that the latent variable is normally distributed has been rejected. This result is coherent with the one of \cite{cagnone2012factor} and with the simulation study, where both the $GH_{T1}$ and $LR_1$ tests show good performance in terms of power to detect non-normality of the latent variable distribution with many items and large sample sizes. However, the result of $GH_{T1}$ is more reliable than $LR_1$ because $GH_{T1}$ never has inflated Type I error rates with large sample sizes.
In this data analysis, we did not consider $GH_{T2}$ and $LR_2$ tests. This decision was made because both tests already reject the null hypothesis when considering only $L=1$, and there is also the issue of the initial values used in the optimization process. 
\section{Discussion}
In this work, we have extended the use of the GH test to detect non-normality of the latent variable distribution in unidimensional IRT model for binary data. The GH test has been obtained by comparing the PL estimator of the 2PL IRT model for binary data with the quasi-ML estimator of the SNP-IRT model, which allows for a more flexible shape of the latent variable distribution. A simpler version than the GH test, referred to as the $GH_T$ test, has been employed because it does not require the inversion of the covariance matrix. Its approximated distribution has been derived using the moment matching method. Two versions of  $GH_T$ have been considered.
First, we evaluated the performance of the $GH_{T1}$ test based on the $SNP_1$ model through a simulation study and real data analysis. We have compared the performance of the $GH_{T1}$ test with the $M_2$ and the $LR_1$ tests and computed  information criteria, such as AIC, BIC, and HQ.
The simulation study has shown that the $GH_{T1}$ test has good performance in terms of Type I error rates under most conditions.  It also exhibits the highest power to detect non-normality of the latent variable distribution when the true latent variable followed a mixture of normals. The $LR_1$ test has inflated or deflated Type I error rates under certain conditions and consistently lower power compared to the $GH_{T1}$ test. Considering that the SNP density when $L=1$ can only capture bimodal and slightly skew-normal distributions, we have adopted the SNP density with $L=2$ in the context of skew-normal distributions. This particular parameterization enables the representation of highly skew-normal distributions for specific combinations of parameter values.
For this reason, we have employed the $GH_{T2}$ test based on the $SNP_2$ model, along with the $LR_2$ test. The $GH_{T1}$ and $GH_{T2}$ tests perform similarly under the skew-normal scenarios and exhibit high power with large sample sizes. The $LR_2$ test has the highest power. However, both the $GH_{T2}$ and $LR_2$ tests required the estimation of the $SNP_2$ model. This poses a challenge since accurate initial values for the true parameter values are required in the optimization process to attain a reliable approximation of the true latent variable and minimize bias in parameter estimates compared to $SNP_0$. As a result, the application of the $SNP_2$ model in the analysis of real data becomes impractical due to these demanding requirements. The $M_2$ test has good performance in terms of Type I error rates but very low or no power to detect the non-normality of the latent variable. Similar results on the low power of the $M_2$ test to detect non-normality of the latent variable distribution have been found by \cite{paek2019detection}. Also \cite{ranger2020analyzing} have found that the $M_2$ test has very low or no power when the misspecification was in the form of an upper boundary item characteristic function. From the simulations it has not been possible to identify the information criterion that has the best performance both under normality and non-normality of the latent variable. AIC tends to select the more complex model in the highest percentages of cases, regardless of the underlying distribution of the latent variable. This is because it tends to favor models with more parameters, which may lead to overfitting.
On the other hand, BIC has the best performance under normality of the latent variable, as it penalizes the number of parameters more strongly than AIC, which makes it more suitable for selecting parsimonious models. However, BIC has the worst performance under non-normality of the latent variable for small sample sizes, as it selects overly simple models that do not capture the complexity of the data. The performance of the HQ criterion is in between the one of AIC and BIC. The results from the real data analysis confirmed the findings from the simulation study, where the $GH_{T1}$ test proved to be very useful in detecting non-normality of the latent variable when the information criteria yielded contradictory results. Overall, the $GH_{T1}$ test emerged as the most powerful tool for detecting non-normality of the latent variable distribution.

Further studies could focus on addressing the issue of initial values in the SNP estimation when $L>1$, making it more applicable in practical contexts. The performance of the GH test implemented with higher-order polynomials could also be evaluated by means of simulations and in real data analysis. Increasing the degree of the polynomial allows for more flexibility in modeling the shape of the latent variable distribution and can enhance both information criteria and test performance.

In addition, the performance of the GH
test in the IRT context could include other types of model violations, as local
dependence or violation of the item characteristic function. In these cases, other types of estimation methods consistent under model misspecification should be considered in order to apply the test.
\bibliographystyle{apalike}
\bibliography{main}
\appendix

\appendix
\section{The mean and variance of the SNP latent variable}
To compute the final estimator $\hat{\boldsymbol{\alpha}}_0$ in formula (14) and $\hat{\boldsymbol{\alpha}}_1$ in (15), it is necessary to compute $\tilde{E}(Z)$ and $\tilde{V}(Z)$ for the latent variable with density in (2). These quantities can be derived analytically.

These quantity are derived for $L=2$. After the optimization process,  $P_L(z)=a_0+a_1z+a_2z^2$ and $h(z|\hat{\boldsymbol{\varphi}})=P_L^2(z)\phi(z)$, where $a_0=sin\hat{\varphi}_1-\frac{1}{\sqrt2}cos\hat{\varphi}_1cos\hat{\varphi}_2$, $a_1=cos\hat{\varphi}_1sin\hat{\varphi}_2$, $a_2=\frac{1}{\sqrt2}cos\hat{\varphi}_1cos\hat{\varphi}_2$.

From \citet{zhang2001linear}
\begin{equation}
    \tilde{E}(Z)=a'M^*a
\end{equation}
where the element in the $i$-th row and $j$-th column of
$M^*$ is $E(z^{i+j-1})$ and $z\sim N(0,1)$. The matrix $M^*$ includes the moment of a standard normal distribution.

When $L=2$
\begin{equation}
    M^*=E\begin{pmatrix}
z & z^2 & z^3\\
z^2 & z^3 & z^4\\
z^3 & z^4 & z^5\\
\end{pmatrix}=\begin{pmatrix}
0 & 1 & 0\\
1 & 0 & 3\\
0 & 3 &0\\
\end{pmatrix}
\end{equation}
and
\begin{equation}
\small
\begin{split}
   \tilde{E}(Z)&=\begin{pmatrix}
sin\hat{\varphi}_1-\frac{1}{\sqrt2}cos\hat{\varphi}_1cos\hat{\varphi}_2& cos\hat{\varphi}_1sin\hat{\varphi}_2 & \frac{1}{\sqrt2}cos\hat{\varphi}_1cos\hat{\varphi}_2
\end{pmatrix}\begin{pmatrix}
0 & 1 & 0\\
1 & 0 & 3\\
0 & 3 &0\\
\end{pmatrix}\begin{pmatrix}
sin\hat{\varphi}_1-\frac{1}{\sqrt2}cos\hat{\varphi}_1cos\hat{\varphi}_2\\ cos\hat{\varphi}_1sin\hat{\varphi}_2 \\ \frac{1}{\sqrt2}cos\hat{\varphi}_1cos\hat{\varphi}_2
\end{pmatrix}=\\&=\begin{pmatrix}
cos\hat{\varphi}_1sin\hat{\varphi}_2& sin\hat{\varphi}_1+\frac{2}{\sqrt2}cos\hat{\varphi}_1cos\hat{\varphi}_2 & 3cos\hat{\varphi}_1sin\hat{\varphi}_2
\end{pmatrix}\begin{pmatrix}
sin\hat{\varphi}_1-\frac{1}{\sqrt2}cos\hat{\varphi}_1cos\hat{\varphi}_2\\ cos\hat{\varphi}_1sin\hat{\varphi}_2 \\ \frac{1}{\sqrt2}cos\hat{\varphi}_1cos\hat{\varphi}_2
\end{pmatrix}=\\
&=2sin\hat{\varphi}_1cos\hat{\varphi}_1sin\hat{\varphi}_2+\frac{4}{\sqrt2}{cos\hat{\varphi}_1}^2cos\hat{\varphi}_2sin\hat{\varphi}_2
\end{split} \label{edc}
\end{equation}

To compute $\tilde{V}(Z)$ we need also $\tilde{E}(Z^2)$. It can be computed as as $a'M^{**}a$, where the element in the $i$-th row and $j$-th column of
$M^{**}$ is $E(z^{i+j})$, and $z \sim N(0,1)$ ( \citealp{zhang2001linear}).
When $L=2$ 
\begin{equation}
    M^{**}=E\begin{pmatrix}
z^2 & z^3 & z^4\\
z^3 & z^4 & z^5\\
z^4 & z^5 & z^6\\
\end{pmatrix}=\begin{pmatrix}
1 & 0 & 3\\
0 & 3 & 0\\
3 & 0 &15\\
\end{pmatrix}
\end{equation}

and 
\begin{equation}
\small
\begin{split}
    \tilde{E}(Z^2)&=\begin{pmatrix}
sin\hat{\varphi}1-\frac{1}{\sqrt2}cos\hat{\varphi}_1cos\hat{\varphi}_2& cos\hat{\varphi}_1sin\hat{\varphi}_2 & \frac{1}{\sqrt2}cos\hat{\varphi}_1cos\hat{\varphi}_2
\end{pmatrix}\begin{pmatrix}
1 & 0 & 3\\
0 & 3 & 0\\
3 & 0 &15\\
\end{pmatrix}\begin{pmatrix}
sin\hat{\varphi}_1-\frac{1}{\sqrt2}cos\hat{\varphi}_1cos\hat{\varphi}_2\\ cos\hat{\varphi}_1sin\hat{\varphi}_2 \\ \frac{1}{\sqrt2}cos\hat{\varphi}_1cos\hat{\varphi}_2
\end{pmatrix}=\\&=\begin{pmatrix}
sin\hat{\varphi}_1+\frac{2}{\sqrt2}cos\hat{\varphi}_1cos\hat{\varphi}_2 & 3
cos\hat{\varphi}_1sin\hat{\varphi}_2 & 3sin\hat{\varphi}_1+\frac{12}{\sqrt2}cos\hat{\varphi}_1cos\hat{\varphi}_2
\end{pmatrix}\begin{pmatrix}
sin\hat{\varphi}_1-\frac{1}{\sqrt2}cos\hat{\varphi}_1cos\hat{\varphi}_2\\ cos\hat{\varphi}_1sin\hat{\varphi}_2 \\ \frac{1}{\sqrt2}cos\hat{\varphi}_1cos\hat{\varphi}_2
\end{pmatrix}=\\
&=sin{\hat{\varphi}_1}^2+\frac{4}{\sqrt2}cos\hat{\varphi}_1sin\hat{\varphi}_1cos\hat{\varphi}_2+3{cos\hat{\varphi}_1}^2sin{\hat{\varphi}_2}^2+5{cos\hat{\varphi}_1}^2cos{\hat{\varphi}_2}^2
\end{split}
\end{equation}
The variance of the latent variable with a SNP density is computed as $\tilde{V}(Z)=\tilde{E}(Z^2)-\tilde{E}(Z)^2$, and we obtain

\begin{equation}
\begin{split}
  \tilde{V}(Z)&=  sin{\hat{\varphi}_1}^2+\frac{4}{\sqrt2}cos\hat{\varphi}_1sin\hat{\varphi}1cos\hat{\varphi}_2+3cos{\hat{\varphi}_1}^2sin{\hat{\varphi}_2}^2+5cos{\hat{\varphi}_1}^2cos{\hat{\varphi}_2}^2-\\&-(2sin\hat{\varphi}_1cos\hat{\varphi}_1sin\hat{\varphi}_2+\frac{4}{\sqrt2}cos{\hat{\varphi}_1}^2cos\hat{\varphi}_2sin\hat{\varphi}_2)^2 \label{vardc}
  \end{split}
\end{equation}
To get the mean and the variance of the latent variable for the $SNP_1$ model,  $\hat{\varphi}_2$ should be set equal to $\frac{\pi}{2}$in equations (\ref{edc}) and (\ref{vardc}).
\section{Additional simulation results}
\subsection{Parameter bias for the $\boldsymbol{SNP_0}$, $\boldsymbol{SNP_1}$ and $\boldsymbol{SNP_2}$ models}
In this section, we report the bias observed in the quasi-ML estimates of the $SNP_1$ model and both full-ML and PL estimates of the $SNP_0$ under scenario C. Additionally, for scenario E, we also report the bias of the quasi-ML estimates of the $SNP_2$ model.
In each scenario considered, the mean bias of each model parameter, indicated with $\theta$, is computed as: $$Bias_{\hat{\theta}}=\frac{\sum_{l=1}^N|\hat{\theta}_l-{\theta}_0|}{N},$$
where $\theta_0$ is the true parameter, $\hat{\theta}_l$ is the estimate of the parameter $\theta$ in the $l$-th replication and $N$ is the number of replications, equal to 500.

Table S1 presents the bias of the parameter estimates for the $SNP_1$ and $SNP_0$ models under scenario $C$,
$p=10$, $n=5000$.

\begin{table}[H] \caption{Bias of the parameter estimates for the $SNP_1$ and $SNP_0$ models  under scenario $C$,
$p=10$, $n=5000$}  
\centering  
\relsize{-1}
\begin{tabular}{c c ccccccccccccccccccccc}  
\hline\hline 
$\theta$&&$SNP_1$(quasi-ML)&$SNP_0$(full-ML)&$SNP_0$(PL) \\
\hline   
$\alpha_{0_1}$&&0.13&0.45&0.71\\
$\alpha_{0_2}$&& 0.06& 0.16&0.30\\
$\alpha_{0_3}$&& 0.05& 0.16& 0.24\\
$\alpha_{0_4}$&&0.21&0.83&1.52\\
$\alpha_{0_5}$&&0.06&0.18&0.28 \\
$\alpha_{0_6}$&&0.15&0.54&1.08\\
$\alpha_{0_7}$&&0.06&0.19&0.28\\
$\alpha_{0_8}$&&0.15&0.55&0.87\\
$\alpha_{0_9}$&&0.04&0.10& 0.18\\
$\alpha_{0_{10}}$&&0.03&0.04&0.06\\
$\alpha_{1_1}$&&0.15&0.52&0.74\\
$\alpha_{1_2}$&&0.08&0.28& 0.43\\
$\alpha_{1_3}$&&0.07&0.15&0.20\\
$\alpha_{1_4}$&&0.24& 0.96& 1.57\\
$\alpha_{1_5}$&&0.08& 0.25& 0.36\\

$\alpha_{1_6}$&&0.18&0.70&1.21\\
$\alpha_{1_7}$& &0.08&0.20&0.28\\
$\alpha_{1_8}$&&0.17&0.61&0.88\\
$\alpha_{1_9}$&&0.06&0.19&0.30\\
$\alpha_{1_{10}}$&&0.04&0.07& 0.13\\
\hline\hline
\end{tabular}\\
\end{table}

Table S2 presents the bias of the parameter estimates for the $SNP_2$, $SNP_1$ and $SNP_0$ models under scenario $E$,
$p=10$, $n=1000$.

\begin{table}[H] \caption{Bias of the parameter estimates for the $SNP_2$, $SNP_1$ and $SNP_0$ models under scenario $E$,
$p=10$, $n=1000$}  
\centering  
\relsize{-1}
\begin{tabular}{c c ccccccccccccccccccccc}  
\hline\hline 
$\theta$&$SNP_2$(quasi-ML)&$SNP_1$(quasi-ML)&$SNP_0$(full-ML)&$SNP_0$(PL) \\
\hline   
$\alpha_{0_1}$&0.09&0.11&0.14&0.20\\
$\alpha_{0_2}$&0.06&0.06&0.06&0.07\\
$\alpha_{0_3}$&0.08&0.10& 0.11&0.14\\
$\alpha_{0_4}$&0.09&0.14& 0.16&0.23\\
$\alpha_{0_5}$&0.06&0.08&0.07&0.09\\
$\alpha_{0_6}$&0.07&0.09&0.09&0.12\\
$\alpha_{0_7}$&0.08&0.09&0.10&0.14\\
$\alpha_{0_8}$&0.09&0.13&0.15&0.23\\
$\alpha_{0_9}$&0.05&0.06&0.05&0.06\\
$\alpha_{0_{10}}$&0.05&0.05&0.05&0.05\\
$\alpha_{1_1}$&0.13&0.19&0.17&0.25\\
$\alpha_{1_2}$&0.08&0.09&0.08&0.07\\
$\alpha_{1_3}$& 0.10&0.13&0.15&0.21\\
$\alpha_{1_4}$&0.15&0.23&0.18&0.24\\
$\alpha_{1_5}$&0.08&0.10&0.09&0.11\\

$\alpha_{1_6}$&0.12 &0.13&0.11&0.11\\
$\alpha_{1_7}$&0.10&0.12&0.14&0.19\\
$\alpha_{1_8}$&0.14&0.23&0.20&0.29\\
$\alpha_{1_9}$&0.08&0.08&0.08&0.08\\
$\alpha_{1_{10}}$&0.07&0.07&0.07&0.07\\
\hline\hline
\end{tabular}\\
\end{table}
\section{Graphs of SNP densities}
In this section, we present some graphs that include the true non-normal latent variable densities used for the simulations, as well as the SNP densities, for scenario C and E. The $\varphi_1$ and $\varphi_2$ parameters in the SNP densities correspond to the median parameter estimates obtained across simulations.

\begin{figure}[H]
 \includegraphics[scale=0.7]{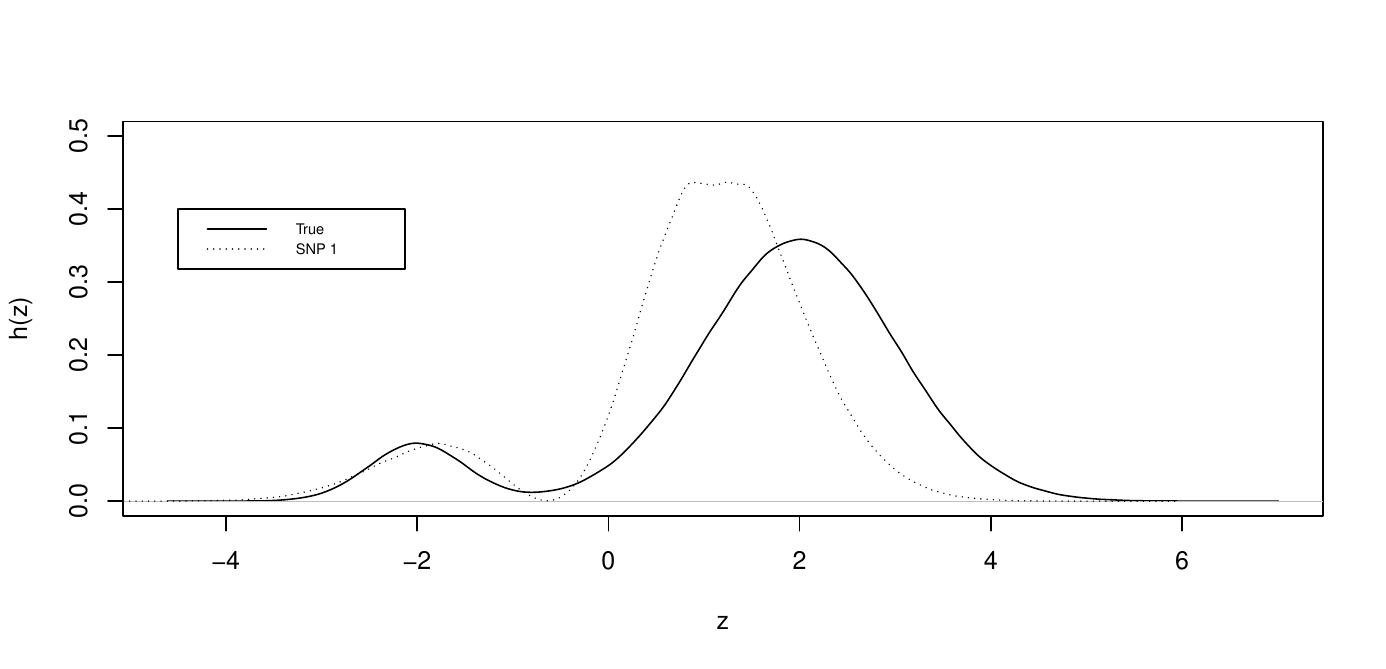}
 \caption{True density and estimated SNP density for scenario C}
\label{table:ta}
 \end{figure}

\begin{figure}[H]
 \includegraphics[scale=0.7]{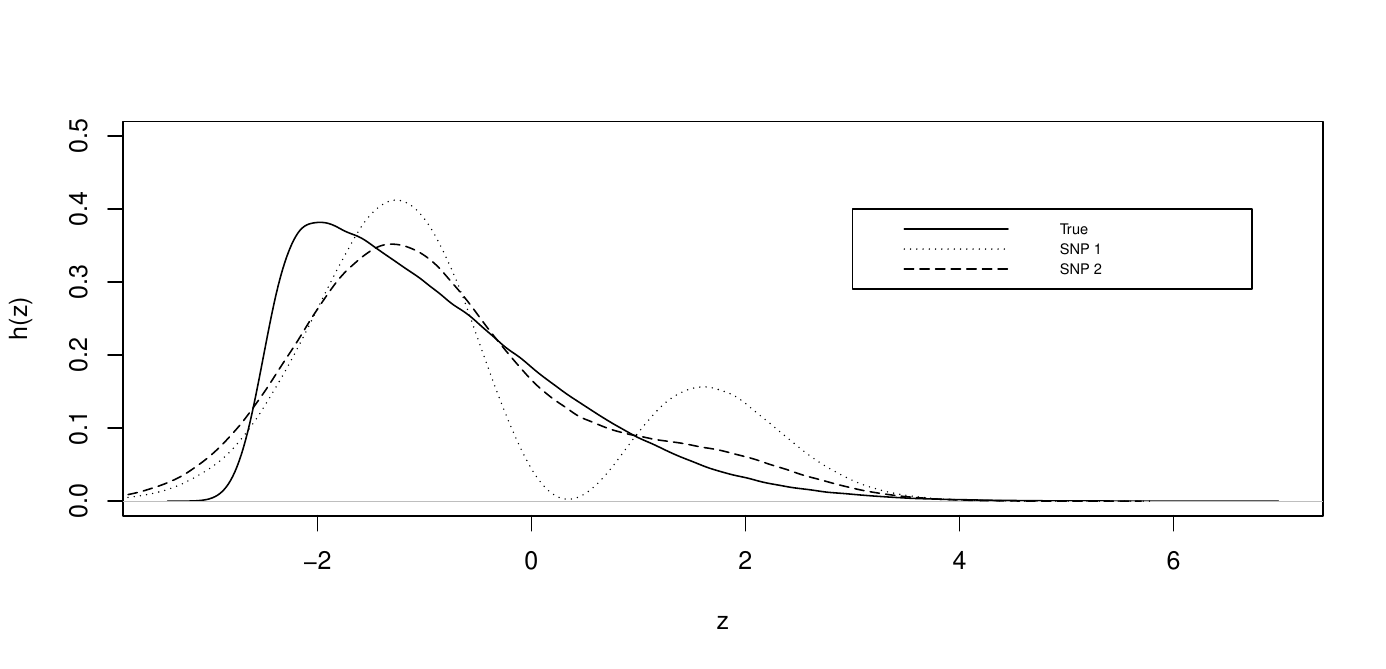}
 \caption{True density
and estimated SNP densities for scenario E}
\label{table:ta3}
 \end{figure}

Figure \ref{table:ta} displays the true density of the latent variable and the estimated SNP density with $L=1$ for scenario C, $p=10$ and $n=5000$.
Figure \ref{table:ta3} illustrates the true density of the latent variable and the estimated SNP densities with $L=1$ and $L=2$ for scenario E, $p=10$ and $n=1000$.

\end{document}